\PassOptionsToPackage{unicode}{hyperref}
\PassOptionsToPackage{hyphens}{url}
\PassOptionsToPackage{dvipsnames,svgnames,x11names}{xcolor}
\documentclass[
]{article}
\usepackage{amsmath,amssymb}
\usepackage{lmodern}
\usepackage{iftex}
\ifPDFTeX
  \usepackage[T1]{fontenc}
  \usepackage[utf8]{inputenc}
  \usepackage{textcomp} 
\else 
  \usepackage{unicode-math}
  \defaultfontfeatures{Scale=MatchLowercase}
  \defaultfontfeatures[\rmfamily]{Ligatures=TeX,Scale=1}
\fi
\IfFileExists{upquote.sty}{\usepackage{upquote}}{}
\IfFileExists{microtype.sty}{
  \usepackage[]{microtype}
  \UseMicrotypeSet[protrusion]{basicmath} 
}{}
\makeatletter
\@ifundefined{KOMAClassName}{
  \IfFileExists{parskip.sty}{%
    \usepackage{parskip}
  }{
    \setlength{\parindent}{0pt}
    \setlength{\parskip}{6pt plus 2pt minus 1pt}}
}{
  \KOMAoptions{parskip=half}}
\makeatother
\usepackage{xcolor}
\providecommand{\tightlist}{%
  \setlength{\itemsep}{0pt}\setlength{\parskip}{0pt}}
\newlength{\cslhangindent}
\newlength{\csllabelwidth}
\newlength{\cslentryspacingunit} 
\newenvironment{CSLReferences}[2] 
 {
  \setlength{\parindent}{0pt}
  \ifodd #1
  \let\oldpar\par
  \def\par{\hangindent=\cslhangindent\oldpar}
  \fi
  \setlength{\parskip}{#2\cslentryspacingunit}
 }%
 {}
\usepackage{calc}

\ifLuaTeX
\usepackage[bidi=basic]{babel}
\else
\usepackage[bidi=default]{babel}
\fi
\babelprovide[main,import]{american}

\def\languageshorthands#1{}
\ifLuaTeX
  \usepackage{selnolig}  
\fi
\IfFileExists{bookmark.sty}{\usepackage{bookmark}}{\usepackage{hyperref}}
\IfFileExists{xurl.sty}{\usepackage{xurl}}{} 
\hypersetup{
  pdftitle={pysersic: A Python package for determining galaxy structural
properties via Bayesian inference, accelerated with jax},
  pdfauthor={Imad Pasha, Tim B. Miller},
  pdflang={en-US},
  colorlinks=true,
  linkcolor={Maroon},
  filecolor={Maroon},
  citecolor={Blue},
  urlcolor={Blue},
  pdfcreator={LaTeX via pandoc}}

\title{pysersic: A Python package for determining galaxy structural
properties via Bayesian inference, accelerated with jax}

\usepackage[affil-it]{authblk}
\usepackage{orcidlink}
\author[1,2%
  *%
  ]{Imad Pasha%
    \,\orcidlink{0000-0002-7075-9931}\,%
    }
\author[1%
  *%
  ]{Tim B. Miller%
    \,\orcidlink{0000-0001-8367-6265}\,%
    }

\affil[1]{Department of Astronomy, Yale University, USA}
\affil[2]{National Science Foundation Graduate Research Fellow}
\affil[*]{These authors contributed equally.}
\date{5 June 2023}

\begin{document}
\maketitle

\hypertarget{summary}{%
\section{Summary}\label{summary}}

The modern standard for measuring structural parameters of galaxies
involves a forward-modeling procedure in which parametric models are fit
directly to images while accounting for the effect of the point-spread
function (PSF). This is an integral step in many extragalactic studies.
The most common parametric form is a Sérsic profile
(\protect\hyperlink{ref-Sersic:1968}{Sersic, 1968}) which is described
by a radial surface brightness profiles following,

\[
I(R) \propto F_{\rm total} \exp \left[\left(\frac{R}{R_e}\right)^{1/n}-1\right],
\]

where the total flux, \(F_{\rm total}\), half-light radius, \(R_e\) and
Sérsic index, \(n\) are the parameters of interest to be fit and
subsequently used to characterize a galaxy's morphology.

Here we present \texttt{pysersic},\footnote{Available via pip or at \url{https://github.com/pysersic/pysersic}.} a Bayesian framework created to
facilitate the inference of structural parameters from galaxy images. It
is written in pure \texttt{Python}, and built using the \texttt{jax}
framework (\protect\hyperlink{ref-Bradbury:2018}{Bradbury et al., 2018})
allowing for just-in-time (JIT) compilation, auto-differentiation and
seamless execution on CPUs, GPUs or TPUs. Inference is performed with
the \texttt{numpyro} (\protect\hyperlink{ref-Bingham:2019}{Bingham et
al., 2019}; \protect\hyperlink{ref-Phan:2019}{Phan et al., 2019})
package utilizing gradient based methods, e.g., No U-Turn Sampling
(NUTS) (\protect\hyperlink{ref-Hoffman:2014}{Hoffman et al., 2014}), for
efficient and robust posterior estimation. \texttt{pysersic} was
designed to have a user-friendly interface, allowing users to fit single
or multiple sources in a few lines of code. It was also designed to
scale to many images, such that it can be seamlessly integrated into
current and future analysis pipelines.

\hypertarget{statement-of-need}{%
\section{Statement of need}\label{statement-of-need}}

Parametric profile fitting has become a ubiquitous and essential tool
for numerous applications including measuring the photometry --- or
total flux --- of galaxies, as well as the investigation of the
structural evolution of galaxies over cosmic time
(\protect\hyperlink{ref-Kawinwanichakij:2021}{Kawinwanichakij et al.,
2021}; \protect\hyperlink{ref-Lange:2015}{Lange et al., 2015};
\protect\hyperlink{ref-Mowla:2019}{Mowla et al., 2019}). This approach
allows one to both extrapolate galaxy surface brightness profiles beyond
the noise limit of images, as well as account for the PSF to accurately
measure the structure of galaxies near the resolution limit of those
images. The empirically derived Sérsic profile is the most common
parametric form for the surface-brightness profile as it provides a
reasonable approximation to nearly all galaxies, given the additional
freedom of the Sérsic index, \(n\), over fixed-index profiles.

Given the long history of Sérsic fitting codes with many available
tools, the development of \texttt{pysersic} was largely motivated by two
related factors, first and foremost of which was the desire to implement
Sérsic fitting in a fully Bayesian context \emph{at speed}. The
\emph{ability} to place the typical Sérsic fitting problem into a
Bayesian context with runtimes that are not prohibitive (the traditional
drawback of MCMC methods) has recently been unlocked by the second
motivation: to leverage the \texttt{jax} library. \texttt{jax} utilizes
JIT compilation to decrease computational runtimes, provides seamless
integration with hardware accelerators such as GPUs and TPUs for further
improvements in performance, and enables automatic differentiation,
facilitating gradient based optimization and sampling methods. Together,
these features greatly increase speed and efficiency, especially when
sampling or optimizing a large number of parameters.

Inference in \texttt{pysersic} is implemented using the \texttt{numpyro}
probabilistic programming language (PPL). This allows for total control
over the priors and methods used for inference. The \texttt{numpyro}
package utilizes \texttt{jax}'s auto-differentiation capabilities for
gradient based samplers such as Hamiltonian Monte Carlo (HMC) and
No-U-Turn-Sampling (NUTS). In addition, there are recently-developed
techniques for posterior estimation, including variational inference
(\protect\hyperlink{ref-Ranganath:2014}{Ranganath et al., 2014})
utilizing normalizing flows (\protect\hyperlink{ref-DeCao:2020}{De Cao
et al., 2020}). These techniques dramatically reduce the number of
likelihood calls required to provide accurate estimates of the posterior
relative to gradient-free methods. Combined with the \texttt{jax}'s JIT
compilation, posteriors can now be generated in a few minutes or less on
modern laptops.

\hypertarget{code-description}{%
\section{Code Description}\label{code-description}}

\texttt{pysersic} was designed to have a user-friendly API with sensible
defaults. Tools are provided to automatically generate priors for all
free parameters based on an initial characterization of a given image
--- but can also easily be set manually. We provide default inference
routines for NUTS MCMC and variational inference using neural flows.
Users can access the underlying \texttt{numpyro} model if desired, to
perform inference using any tools available within the \texttt{numpyro}
ecosystem. The goal for \texttt{pysersic} is to provide a reasonable
defaults for new users interested in a handful of galaxies, yet maintain
the ability for advanced users to tweak options as necessary to perform
inference for entire surveys.

A crucial component of any Sérsic fitting code is a efficient and
accurate rendering algorithm. Sérsic profiles with high index,
\(n\gtrsim 3\) are notoriously difficult to render accurately given the
steep increase in brightness as \(r \rightarrow 0\). In
\texttt{pysersic}, the \texttt{rendering} module is kept separate from
the frontend API and inference modules, such that different algorithms
can be interchanged and therefore easily tested (and hopefully encourage
innovation as well). In this initial release, we provide three
algorithms. The first is a traditional rendering algorithm in which the
intrinsic profile is rendered in real space, with oversampling in the
center to ensure accurate results for high index profiles. The second
and third methods render the profiles in Fourier space, providing
accurate results even for strongly peaked profiles and avoiding
artifacts due to pixelization. In \texttt{pysersic}, this is achieved by
representing the profiles using a series of Gaussian following the
algorithm presented in Shajib
(\protect\hyperlink{ref-Shajib:2019}{2019}). We include one algorithm
that is fully based in Fourier space, along with a version of the hybrid
real-Fourier algorithm introduced in Lang
(\protect\hyperlink{ref-Lang:2020}{2020}) which helps avoid some of the
aliasing present when rendering solely in Fourier space.

\hypertarget{related-software}{%
\section{Related Software}\label{related-software}}

There is a long history and many software tools designed for Sérsic
profile fitting. Some of the most popular libraries are listed below.

\begin{itemize}
\tightlist
\item
  \texttt{galfit} (\protect\hyperlink{ref-Peng:2002}{Peng et al., 2002})
\item
  \texttt{imfit} (\protect\hyperlink{ref-Erwin:2015}{Erwin, 2015})
\item
  \texttt{profit} (\protect\hyperlink{ref-Robotham:2017}{Robotham et
  al., 2017})
\item
  \texttt{galight} (\protect\hyperlink{ref-Ding:2021}{Ding et al.,
  2021}), which is built on top of \texttt{lenstronomy}
  (\protect\hyperlink{ref-Birrer:2021}{Birrer et al., 2021})
\item
  \texttt{PetroFit} (\protect\hyperlink{ref-Geda:2022}{Geda et al.,
  2022})
\item
  \texttt{PyAutoGalaxy}
  (\protect\hyperlink{ref-Nightingale:2023}{Nightingale et al., 2023})
\end{itemize}

\hypertarget{software-citations}{%
\section{Software Citations}\label{software-citations}}

\texttt{pysersic} makes use of the following packages:

\begin{itemize}
\tightlist
\item
  arviz (\protect\hyperlink{ref-arviz:2019}{Kumar et al., 2019})
\item
  asdf (\protect\hyperlink{ref-asdf}{D'Avella et al., 2023})
\item
  astropy (\protect\hyperlink{ref-astropy:2013}{Astropy Collaboration et
  al., 2013}, \protect\hyperlink{ref-astropy:2018}{2018},
  \protect\hyperlink{ref-astropy:2022}{2022})
\item
  corner (\protect\hyperlink{ref-corner}{Foreman-Mackey, 2016})
\item
  jax (\protect\hyperlink{ref-Bradbury:2018}{Bradbury et al., 2018})
\item
  matplotlib (\protect\hyperlink{ref-Hunter:2007}{Hunter, 2007})
\item
  numpy (\protect\hyperlink{ref-Harris:2020}{Harris et al., 2020})
\item
  numpyro (\protect\hyperlink{ref-Bingham:2019}{Bingham et al., 2019};
  \protect\hyperlink{ref-Phan:2019}{Phan et al., 2019})
\item
  pandas (\protect\hyperlink{ref-reback:2020}{team, 2020})
\item
  photutils (\protect\hyperlink{ref-Bradley:2022}{Bradley et al., 2022})
\item
  pytest (\protect\hyperlink{ref-pytest}{Krekel et al., 2004})
\item
  scipy (\protect\hyperlink{ref-Scipy:2020}{Virtanen et al., 2020})
\item
  tqdm (\protect\hyperlink{ref-tqdm}{Costa-Luis et al., 2023})
\end{itemize}

\hypertarget{acknowledgements}{%
\section{Acknowledgements}\label{acknowledgements}}

We acknowledge Pieter van Dokkum for useful conversations surrounding
the design and implementation of \texttt{pysersic}.

\hypertarget{references}{%
\section*{References}\label{references}}
\addcontentsline{toc}{section}{References}

\hypertarget{refs}{}
\begin{CSLReferences}{1}{0}
\leavevmode\vadjust pre{\hypertarget{ref-astropy:2022}{}}%
Astropy Collaboration, Price-Whelan, A. M., Lim, P. L., Earl, N.,
Starkman, N., Bradley, L., Shupe, D. L., Patil, A. A., Corrales, L.,
Brasseur, C. E., N"othe, M., Donath, A., Tollerud, E., Morris, B. M.,
Ginsburg, A., Vaher, E., Weaver, B. A., Tocknell, J., Jamieson, W.,
\ldots{} Astropy Project Contributors. (2022). {The Astropy Project:
Sustaining and Growing a Community-oriented Open-source Project and the
Latest Major Release (v5.0) of the Core Package}. \emph{Apj},
\emph{935}(2), 167. \url{https://doi.org/10.3847/1538-4357/ac7c74}

\leavevmode\vadjust pre{\hypertarget{ref-astropy:2018}{}}%
Astropy Collaboration, Price-Whelan, A. M., Sipőcz, B. M., Günther, H.
M., Lim, P. L., Crawford, S. M., Conseil, S., Shupe, D. L., Craig, M.
W., Dencheva, N., Ginsburg, A., Vand erPlas, J. T., Bradley, L. D.,
Pérez-Suárez, D., de Val-Borro, M., Aldcroft, T. L., Cruz, K. L.,
Robitaille, T. P., Tollerud, E. J., \ldots{} Astropy Contributors.
(2018). {The Astropy Project: Building an Open-science Project and
Status of the v2.0 Core Package}. \emph{156}(3), 123.
\url{https://doi.org/10.3847/1538-3881/aabc4f}

\leavevmode\vadjust pre{\hypertarget{ref-astropy:2013}{}}%
Astropy Collaboration, Robitaille, T. P., Tollerud, E. J., Greenfield,
P., Droettboom, M., Bray, E., Aldcroft, T., Davis, M., Ginsburg, A.,
Price-Whelan, A. M., Kerzendorf, W. E., Conley, A., Crighton, N.,
Barbary, K., Muna, D., Ferguson, H., Grollier, F., Parikh, M. M., Nair,
P. H., \ldots{} Streicher, O. (2013). {Astropy: A community Python
package for astronomy}. \emph{558}, A33.
\url{https://doi.org/10.1051/0004-6361/201322068}

\leavevmode\vadjust pre{\hypertarget{ref-Bingham:2019}{}}%
Bingham, E., Chen, J. P., Jankowiak, M., Obermeyer, F., Pradhan, N.,
Karaletsos, T., Singh, R., Szerlip, P. A., Horsfall, P., \& Goodman, N.
D. (2019). Pyro: Deep universal probabilistic programming. \emph{J.
Mach. Learn. Res.}, \emph{20}, 28:1--28:6.
\url{http://jmlr.org/papers/v20/18-403.html}

\leavevmode\vadjust pre{\hypertarget{ref-Birrer:2021}{}}%
Birrer, S., Shajib, A. J., Gilman, D., Galan, A., Aalbers, J., Millon,
M., Morgan, R., Pagano, G., Park, J. W., Teodori, L., Tessore, N.,
Ueland, M., Vyvere, L. V. de, Wagner-Carena, S., Wempe, E., Yang, L.,
Ding, X., Schmidt, T., Sluse, D., \ldots{} Amara, A. (2021). Lenstronomy
II: A gravitational lensingW software ecosystem. \emph{Journal of Open
Source Software}, \emph{6}(62), 3283.
\url{https://doi.org/10.21105/joss.03283}

\leavevmode\vadjust pre{\hypertarget{ref-Bradbury:2018}{}}%
Bradbury, J., Frostig, R., Hawkins, P., Johnson, M. J., Leary, C.,
Maclaurin, D., Necula, G., Paszke, A., VanderPlas, J., Wanderman-Milne,
S., \& Zhang, Q. (2018). \emph{{JAX}: Composable transformations of
{P}ython+{N}um{P}y programs} (Version 0.3.13).
\url{http://github.com/google/jax}

\leavevmode\vadjust pre{\hypertarget{ref-Bradley:2022}{}}%
Bradley, L., Sipőcz, B., Robitaille, T., Tollerud, E., Vinícius, Z.,
Deil, C., Barbary, K., Wilson, T. J., Busko, I., Donath, A., Günther, H.
M., Cara, M., Lim, P. L., Meßlinger, S., Conseil, S., Bostroem, A.,
Droettboom, M., Bray, E. M., Bratholm, L. A., \ldots{} Souchereau, H.
(2022). \emph{Astropy/photutils: 1.5.0} (Version 1.5.0). Zenodo.
\url{https://doi.org/10.5281/zenodo.6825092}

\leavevmode\vadjust pre{\hypertarget{ref-tqdm}{}}%
Costa-Luis, C. da, Larroque, S. K., Altendorf, K., Mary, H.,
richardsheridan, Korobov, M., Raphael, N., Ivanov, I., Bargull, M.,
Rodrigues, N., Chen, G., Lee, A., Newey, C., CrazyPython, JC, Zugnoni,
M., Pagel, M. D., mjstevens777, Dektyarev, M., \ldots{} Nordlund, M.
(2023). \emph{{tqdm: A fast, Extensible Progress Bar for Python and
CLI}} (Version v4.65.0). Zenodo.
\url{https://doi.org/10.5281/zenodo.7697295}

\leavevmode\vadjust pre{\hypertarget{ref-asdf}{}}%
D'Avella, D., Jamieson, W., Droettboom, M., Slavich, E., Graham, B.,
Robitaille, T., Dencheva, N., perrygreenfield, Simon, B., MacDonald, K.,
Bray, E. M., Burnett, Z., Davies, J., Mumford, S., Markovtsev, V.,
Tollerud, E., Sipőcz, B., Bradley, L., Fabry, Ç., \ldots{} Ginsburg, A.
(2023). \emph{Asdf-format/asdf: 2.15.0} (Version 2.15.0). Zenodo.
\url{https://doi.org/10.5281/zenodo.7799772}

\leavevmode\vadjust pre{\hypertarget{ref-DeCao:2020}{}}%
De Cao, N., Aziz, W., \& Titov, I. (2020). Block neural autoregressive
flow. \emph{Uncertainty in Artificial Intelligence}, 1263--1273.

\leavevmode\vadjust pre{\hypertarget{ref-Ding:2021}{}}%
Ding, X., Birrer, S., Treu, T., \& Silverman, J. D. (2021). {Galaxy
shapes of Light (GaLight): a 2D modeling of galaxy images}. \emph{arXiv
e-Prints}, arXiv:2111.08721.
\url{https://doi.org/10.48550/arXiv.2111.08721}

\leavevmode\vadjust pre{\hypertarget{ref-Erwin:2015}{}}%
Erwin, P. (2015). {IMFIT: A Fast, Flexible New Program for Astronomical
Image Fitting}. \emph{799}(2), 226.
\url{https://doi.org/10.1088/0004-637X/799/2/226}

\leavevmode\vadjust pre{\hypertarget{ref-corner}{}}%
Foreman-Mackey, D. (2016). Corner.py: Scatterplot matrices in python.
\emph{The Journal of Open Source Software}, \emph{1}(2), 24.
\url{https://doi.org/10.21105/joss.00024}

\leavevmode\vadjust pre{\hypertarget{ref-Geda:2022}{}}%
Geda, R., Crawford, S. M., Hunt, L., Bershady, M., Tollerud, E., \&
Randriamampandry, S. (2022). {PetroFit: A Python Package for Computing
Petrosian Radii and Fitting Galaxy Light Profiles}. \emph{163}(5), 202.
\url{https://doi.org/10.3847/1538-3881/ac5908}

\leavevmode\vadjust pre{\hypertarget{ref-Harris:2020}{}}%
Harris, C. R., Millman, K. J., Walt, S. J. van der, Gommers, R.,
Virtanen, P., Cournapeau, D., Wieser, E., Taylor, J., Berg, S., Smith,
N. J., Kern, R., Picus, M., Hoyer, S., Kerkwijk, M. H. van, Brett, M.,
Haldane, A., Río, J. F. del, Wiebe, M., Peterson, P., \ldots{} Oliphant,
T. E. (2020). Array programming with {NumPy}. \emph{Nature},
\emph{585}(7825), 357--362.
\url{https://doi.org/10.1038/s41586-020-2649-2}

\leavevmode\vadjust pre{\hypertarget{ref-Hoffman:2014}{}}%
Hoffman, M. D., Gelman, A., \& others. (2014). The no-u-turn sampler:
Adaptively setting path lengths in hamiltonian monte carlo. \emph{J.
Mach. Learn. Res.}, \emph{15}(1), 1593--1623.

\leavevmode\vadjust pre{\hypertarget{ref-Hunter:2007}{}}%
Hunter, J. D. (2007). Matplotlib: A 2D graphics environment.
\emph{Computing in Science \& Engineering}, \emph{9}(3), 90--95.
\url{https://doi.org/10.1109/MCSE.2007.55}

\leavevmode\vadjust pre{\hypertarget{ref-Kawinwanichakij:2021}{}}%
Kawinwanichakij, L., Silverman, J. D., Ding, X., George, A., Damjanov,
I., Sawicki, M., Tanaka, M., Taranu, D. S., Birrer, S., Huang, S., Li,
J., Onodera, M., Shibuya, T., \& Yasuda, N. (2021). {Hyper Suprime-Cam
Subaru Strategic Program: A Mass-dependent Slope of the Galaxy Size-Mass
Relation at z \textless{} 1}. \emph{921}(1), 38.
\url{https://doi.org/10.3847/1538-4357/ac1f21}

\leavevmode\vadjust pre{\hypertarget{ref-pytest}{}}%
Krekel, H., Oliveira, B., Pfannschmidt, R., Bruynooghe, F., Laugher, B.,
\& Bruhin, F. (2004). \emph{Pytest}.
\url{https://github.com/pytest-dev/pytest}

\leavevmode\vadjust pre{\hypertarget{ref-arviz:2019}{}}%
Kumar, R., Carroll, C., Hartikainen, A., \& Martin, O. (2019). ArviZ a
unified library for exploratory analysis of bayesian models in python.
\emph{Journal of Open Source Software}, \emph{4}(33), 1143.
\url{https://doi.org/10.21105/joss.01143}

\leavevmode\vadjust pre{\hypertarget{ref-Lang:2020}{}}%
Lang, D. (2020). {A hybrid Fourier--Real Gaussian Mixture method for
fast galaxy--PSF convolution}. \emph{arXiv e-Prints}, arXiv:2012.15797.
\url{https://doi.org/10.48550/arXiv.2012.15797}

\leavevmode\vadjust pre{\hypertarget{ref-Lange:2015}{}}%
Lange, R., Driver, S. P., Robotham, A. S. G., Kelvin, L. S., Graham, A.
W., Alpaslan, M., Andrews, S. K., Baldry, I. K., Bamford, S.,
Bland-Hawthorn, J., Brough, S., Cluver, M. E., Conselice, C. J., Davies,
L. J. M., Haeussler, B., Konstantopoulos, I. S., Loveday, J., Moffett,
A. J., Norberg, P., \ldots{} Wilkins, S. M. (2015). {Galaxy And Mass
Assembly (GAMA): mass-size relations of z \textless{} 0.1 galaxies
subdivided by S{é}rsic index, colour and morphology}. \emph{447}(3),
2603--2630. \url{https://doi.org/10.1093/mnras/stu2467}

\leavevmode\vadjust pre{\hypertarget{ref-Mowla:2019}{}}%
Mowla, L. A., van Dokkum, P., Brammer, G. B., Momcheva, I., van der Wel,
A., Whitaker, K., Nelson, E., Bezanson, R., Muzzin, A., Franx, M.,
MacKenty, J., Leja, J., Kriek, M., \& Marchesini, D. (2019).
{COSMOS-DASH: The Evolution of the Galaxy Size-Mass Relation since z
{\(\sim\)} 3 from New Wide-field WFC3 Imaging Combined with
CANDELS/3D-HST}. \emph{880}(1), 57.
\url{https://doi.org/10.3847/1538-4357/ab290a}

\leavevmode\vadjust pre{\hypertarget{ref-Nightingale:2023}{}}%
Nightingale, James. W., Amvrosiadis, A., Hayes, R. G., He, Q.,
Etherington, A., Cao, X., Cole, S., Frawley, J., Frenk, C. S., Lange,
S., Li, R., Massey, R. J., Negrello, M., \& Robertson, A. (2023).
PyAutoGalaxy: Open-source multiwavelength galaxy structure \&
morphology. \emph{Journal of Open Source Software}, \emph{8}(81), 4475.
\url{https://doi.org/10.21105/joss.04475}

\leavevmode\vadjust pre{\hypertarget{ref-Peng:2002}{}}%
Peng, C. Y., Ho, L. C., Impey, C. D., \& Rix, H.-W. (2002). {Detailed
Structural Decomposition of Galaxy Images}. \emph{124}(1), 266--293.
\url{https://doi.org/10.1086/340952}

\leavevmode\vadjust pre{\hypertarget{ref-Phan:2019}{}}%
Phan, D., Pradhan, N., \& Jankowiak, M. (2019). Composable effects for
flexible and accelerated probabilistic programming in NumPyro.
\emph{arXiv Preprint arXiv:1912.11554}.

\leavevmode\vadjust pre{\hypertarget{ref-Ranganath:2014}{}}%
Ranganath, R., Gerrish, S., \& Blei, D. (2014). Black box variational
inference. \emph{Artificial Intelligence and Statistics}, 814--822.

\leavevmode\vadjust pre{\hypertarget{ref-Robotham:2017}{}}%
Robotham, A. S. G., Taranu, D. S., Tobar, R., Moffett, A., \& Driver, S.
P. (2017). {PROFIT: Bayesian profile fitting of galaxy images}.
\emph{466}(2), 1513--1541. \url{https://doi.org/10.1093/mnras/stw3039}

\leavevmode\vadjust pre{\hypertarget{ref-Sersic:1968}{}}%
Sersic, J. L. (1968). \emph{{Atlas de Galaxias Australes}}.

\leavevmode\vadjust pre{\hypertarget{ref-Shajib:2019}{}}%
Shajib, A. J. (2019). {Unified lensing and kinematic analysis for any
elliptical mass profile}. \emph{488}(1), 1387--1400.
\url{https://doi.org/10.1093/mnras/stz1796}

\leavevmode\vadjust pre{\hypertarget{ref-reback:2020}{}}%
team, T. pandas development. (2020). \emph{Pandas-dev/pandas: pandas}
(latest). Zenodo. \url{https://doi.org/10.5281/zenodo.3509134}

\leavevmode\vadjust pre{\hypertarget{ref-Scipy:2020}{}}%
Virtanen, P., Gommers, R., Oliphant, T. E., Haberland, M., Reddy, T.,
Cournapeau, D., Burovski, E., Peterson, P., Weckesser, W., Bright, J.,
van der Walt, S. J., Brett, M., Wilson, J., Millman, K. J., Mayorov, N.,
Nelson, A. R. J., Jones, E., Kern, R., Larson, E., \ldots{} SciPy 1.0
Contributors. (2020). {{SciPy} 1.0: Fundamental Algorithms for
Scientific Computing in Python}. \emph{Nature Methods}, \emph{17},
261--272. \url{https://doi.org/10.1038/s41592-019-0686-2}

\end{CSLReferences}

\end{document}